\documentclass{aa}
\usepackage{epsf}

\newcommand{\La}{\mbox{${\rm Ly\alpha}$}}
\newcommand{\Lb}{\mbox{${\rm Ly\beta}$}}
\newcommand{\Lg}{\mbox{${\rm Ly\gamma}$}}
\newcommand{\Line}[3]{\Ion{#1}{#2}\,$\lambda$\,#3}
\newcommand{\Lines}[3]{\Ion{#1}{#2}\,$\lambda\lambda$\,#3}
\newcommand{\Ion}[2]{#1{\,\scriptsize #2}}
\newcommand{\Nh}{\mbox{$N_{\rm H}$}}
\newcommand{\Nhi}{\mbox{$N_{\rm H_I}$}}
\newcommand{\Rwd}{\mbox{$R_{\rm wd}$}}
\newcommand{\Mwd}{\mbox{$M_{\rm wd}$}}
\newcommand{\Twd}{\mbox{$T_{\rm wd}$}}
\newcommand{\Msec}{\mbox{$M_{\rm sec}$}}

\newcommand{\Tspot}{\mbox{$T_{\rm spot}$}}
\newcommand{\Tcent}{\mbox{$T_{\rm cent}$}}
\newcommand{\Oang}{\mbox{$\theta_{\rm spot}$}}
\newcommand{\Beta}{\mbox{$\beta_{\rm spot}$}}
\renewcommand{\Phi}{\mbox{$\phi_{\rm spot}$}}
\newcommand{\Teff}{\mbox{$T_{\rm eff}$}}
\newcommand{\Porb}{\mbox{$P_{\rm orb}$}}
\newcommand{\pmag}{\mbox{$\phi_{\rm mag}$}}
\newcommand{\porb}{\mbox{$\phi_{\rm orb}$}}
\newcommand{\Msun}{\mbox{$M_{\odot}$}}

\newcommand{\kms}{\mbox{$\rm km\,s^{-1}$}}
\newcommand{\FN}[1]{\settowidth{\width}{$^{#1)}$}\mbox{$^{#1)}$}\hspace*{-\width}}
\newcommand{\fn}[1]{\mbox{$^{#1)}$}}
\newlength{\width}
\newcommand{\draft}[1]{
}
\draft{Revised, 24-July-1998}

\begin{document}

\thesaurus{06      
          (02.01.2 
           02.12.1 
           08.02.1 
           08.09.2 
           08.23.1 
           13.21.5 
          )}
\title{HST/GHRS observations of AM\,Herculis
\thanks{Based on observations made with the NASA/ESA Hubble Space
Telescope, obtained at the Space Telescope Science Institute, which is
operated by the Association of Universities for Research in Astronomy,
Inc., under NASA contract NAS 5-26555.}}

\author{
B.T. G\"ansicke\inst{1},
D.W. Hoard\inst{2},
K. Beuermann\inst{1},
E.M. Sion\inst{3},
P. Szkody\inst{2}
}

\offprints{boris@uni-sw.gwdg.de}

\institute{
  Universit\"ats-Sternwarte, Geismarlandstr. 11, 37083 G\"ottingen, 
  Germany
\and
Department of Astronomy, University of Washington, Seattle, WA 98195, USA
\and
  Department of Astronomy \& Astrophysics,  Villanova 
  University, Villanova, PA 19085, USA.
}

\date{Received \underline{\hskip2cm} / Accepted 6 August 1998}

\authorrunning{G\"ansicke et al.}

\maketitle 

\begin{abstract}
We present phase-resolved spectroscopy of AM\,Herculis
obtained with the HST/GHRS when the system was in a high state.
The ultraviolet light curve shows a quasi-sinusoidal modulation, which
can be explained by a hot spot on the rotating white dwarf.  The broad
\La\ absorption expected for photospheric radiation of a moderately hot
white dwarf is largely filled in with emission.  The UV/FUV spectrum
of AM\,Her in high state can be quantitatively understood by a
two-component model consisting of the unheated white dwarf plus a
blackbody-like radiating hot spot.
A kinematic study of the strong UV emission lines using Doppler
tomography is presented. The characteristics of the low ionization
species lines and the \Ion{Si}{IV} doublet can be explained within the
classical picture, as broad emission from the accretion stream and narrow
emission from the heated hemisphere of the secondary.  However, we
find that the narrow emission of the \Ion{N}{V} doublet originates
from material of low velocity dispersion located somewhere between
$L_1$ and the centre of mass.
The high signal-to-noise spectra contain a multitude of interstellar
absorption lines but no metal absorption lines from the white dwarf
photosphere.

\end{abstract}

\section{Introduction}
In AM\,Herculis stars (see Warner 1995 for a monograph), three main
sources of ultraviolet (UV) emission are present: the white dwarf
photosphere, the illuminated accretion stream, and the heated
secondary star. While the white dwarf contributes mostly in the
continuum, the accretion stream and the heated atmosphere of the
secondary star are sources of strong emission lines. Phase-resolved UV
spectroscopy can, therefore, reveal details of the temperature
structure on the white dwarf surface as well as kinematic information
of the various emission regions within the binary system.

In a previous paper, we have reported phase-resolved UV observations
of AM\,Her, obtained with IUE during both low and high states
(G\"ansicke et al. 1995; hereafter Paper\,1). During both states, a
modulation of the UV continuum flux, peaking at the phase of maximum
hard X-ray flux, was detected. In the low state, this flux modulation
is accompanied by an orbital variation of the broad \La\ absorption
profile. The modulation of both continuum flux and \La\ absorption
width can be explained with a moderately hot spot near the main
accretion pole on the white dwarf. The spot temperatures estimated
from the IUE data were $\simeq24\,000$\,K and $\ga37\,000$\,K in the
low state and the high state, respectively, with the spot covering
$f\sim0.1$ of the white dwarf surface. The unheated regions of the
white dwarf have $T\simeq20\,000$\,K (Heise \& Verbunt 1988;
Paper\,1).
Considering that the sum of the observed hard X-ray flux and cyclotron
emission roughly equals the UV excess flux of the spot, we concluded
that irradiation by emission from the hot post-shock plasma is the
most probable cause for the heating of the spot. A puzzling result
from ORFEUS-I FUV observations of AM\,Her in high state was the
absence of \Lb\ and \Lg\ absorption lines (Raymond et al. 1995). The
poor resolution of the IUE data gave only limited evidence for
the presence of a \La\ absorption line from the white dwarf
photosphere during the high state, so that a full test of the hot
spot-hypothesis had to await dedicated HST observations.

A general spectroscopic characteristic of polars in their high states
are complexly structured emission lines. At least two components, a
broad and a narrow one, can be identified.  The common belief is that
the broad component originates in the stream while the narrow
component arises from the irradiated face of the secondary
(e.g. Liebert \& Stockman 1985). A beautiful example where three
different line components can be discerned and identified with the
secondary star, and the free-fall and magnetic coupled parts of the
accretion stream is HU\,Aqr (Schwope et al. 1997). Doppler tomography
reveals that the narrow line emission from the secondary in HU\,Aqr is
asymmetric , probably due to shielding of the leading hemisphere by
the accretion stream/curtain. In AM\,Her, broad and narrow components
have been detected in various optical emission lines.  Discussion has
been stimulated by the fact that the individual lines differ in their
radial velocity amplitudes: e.g. $\sim75$\,\kms\ in \Ion{He}{II},
$\sim120$\,\kms\ in \Ion{He}{I} (Greenstein et al. 1979) and
$\sim150$\,\kms\ in \Ion{Ca}{I} (Young \& Schneider 1979). Absorption
lines show even larger velocities: $\sim200$\,\kms\ in \Ion{Na}{I}
(Southwell et al. 1995). Hence, the narrow emission lines and the
absorption lines originate on different parts of the secondary star,
and interpreting their radial velocities as the $K_2$ velocity
amplitude of the secondary star is ambiguous.

In this paper, we report the first high-resolution phase-resolved UV
spectra of AM\,Herculis obtained with the Goddard High Resolution
Spectrograph (GHRS) onboard the Hubble Space Telescope (HST).

\section{HST/GHRS Observations\label{hstobs}}
HST/GHRS observations of AM\,Her were carried out in January 1997
during a rare opportunity when the system was located in a continuous
viewing zone of HST (Table\,1). The total exposure time was slightly
longer than the binary orbital period, $\Porb=185.6$\,min. At the time
of the observations, AAVSO observations showed AM\,Her to be in a high
state at $V\approx13.0$.
The GHRS spectra were taken in the ACCUM mode through the 2" Large
Science Aperture (LSA). In order to cover both wings of \La\, the
central wavelength was set to 1292\,\AA\ resulting in a spectral
coverage of 1150$-$1435\,\AA\ with a nominal FWHM resolution of
$\sim0.6$\,\AA\ ($\sim125-150$\,\kms). The spectra were acquired with
a time resolution of 31.4\,sec, corresponding to an orbital phase
resolution of $\Delta\phi=2.82\times10^{-3}$, with a net exposure time
of 27.2\,sec per spectrum. The observation was interrupted for
$\sim5$\,min for a SPYBAL calibration, causing a gap in the phase
coverage of $\Delta\phi=0.039$. A total of 341 spectra were obtained.

\begin{table}[t]
\caption[]{HST/GHRS observations of AM\,Her on Jan\,4, 1997.}
\begin{flushleft}
\begin{tabular}{lcc}
\hline\noalign{\smallskip}
Dataset: & Z3DM0305T & Z3DM0308T \\ 
\hline\noalign{\smallskip}
Exp. start (UT)       & 2:03:45$-$3:56:15 & 4:01:52$-$5:48:45 \\
Exp. time  (sec)      & 5848 & 5413 \\
No. of spectra        & 215  & 199 \\
Phase coverage (\porb)& $0.497-0.103$  & $0.142-0.705$\\
\noalign{\smallskip}\hline
\end{tabular}
\end{flushleft}
\end{table}

Regular monitoring observations with the GHRS/G140L grating have
revealed a slowly decreasing sensitivity below 1200\,\AA\ (Sherbert et
al. 1997). The response at 1150\,\AA\ was reduced in late 1997 by 15\%
with respect to the sensitivity just after the Service Mission 1 in
December 1993.  We therefore recalibrated our GHRS data of AM\,Her
with the \verb\calhrs\ routine of \verb\stsdas\, using the
time-dependent flux calibration data given by Sherbert \& Hulbert
(1997). We caution, however, that the absolute fluxes at the very blue
end of the spectra ($\lambda\la1180$\,\AA) may be still somewhat on
the low side (Sect.\,\ref{s-lcfit}).

The mid-exposure times of the individual GHRS spectra were converted
into magnetic orbital phases \pmag\ using the ephemeris determined by
Tapia (see Heise \& Verbunt 1988). The magnetic phases were
subsequently converted into conventional binary orbital phases \porb\
via the relation
\begin{equation} 
\porb = \pmag + 0.367.
\end{equation} 
The offset  was obtained by comparison to the zero point in the optical
ephemeris determined by Martin (1988; see also Southwell et al.
1995): $\porb=0.0$ corresponds to the inferior conjunction of the
secondary. 
Note that the period given by Heise \& Verbunt (1988) has a smaller
error than that given by Southwell et al. (1995) and the former should
preferably be used.
Even though \pmag\ is historically used in the analysis of UV and
X-ray light curves of AM\,Her, we adopt \porb\ as the natural choice
for the discussion of the geometry within the binary system. For
convenience, light curves and radial velocity curves will be also
labelled with \pmag.

\section{Analysis and Results}
\begin{figure*}
\begin{center}
\mbox{\epsfxsize=18cm\epsfbox{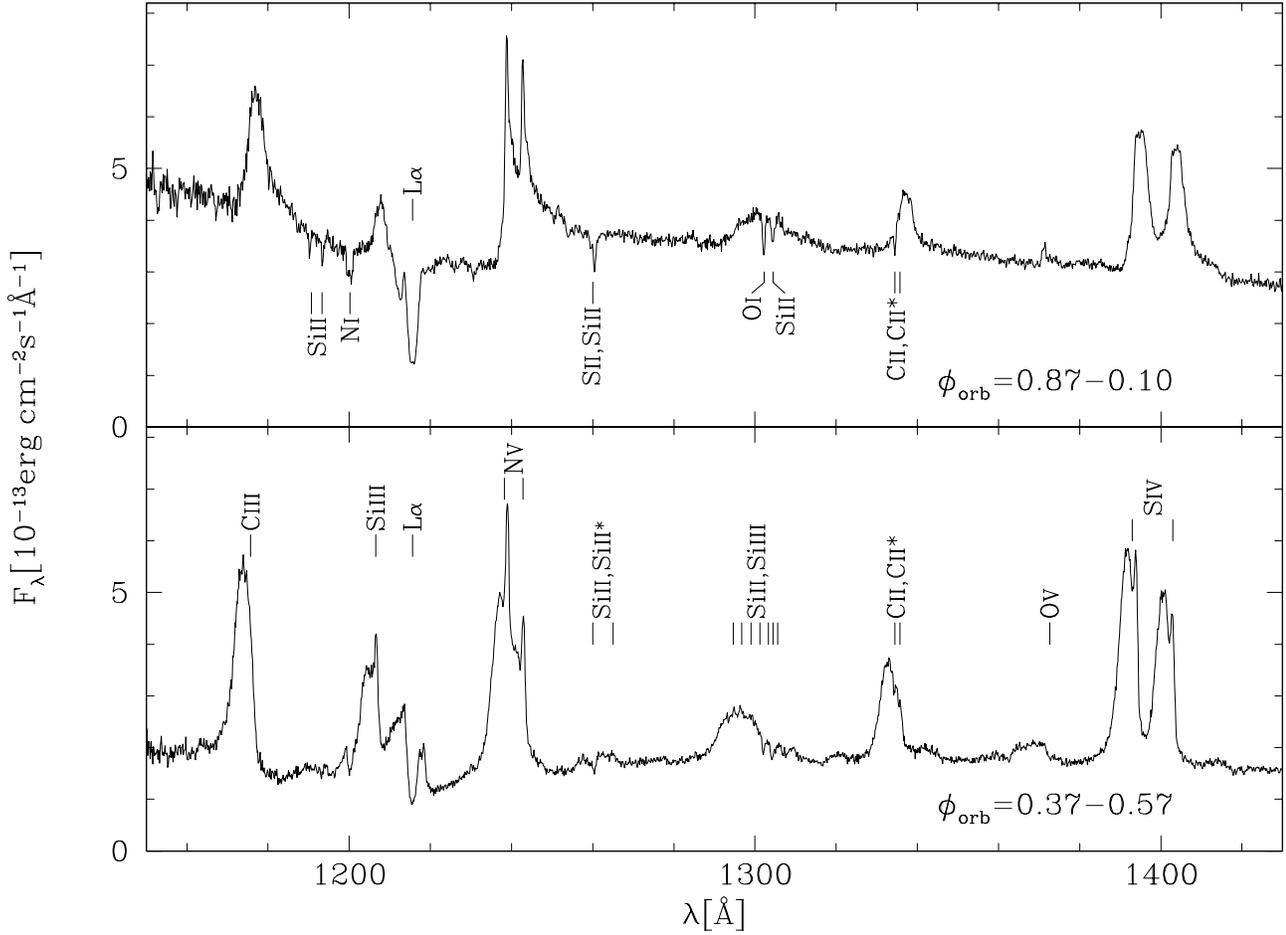}}
\caption[]{\label{ave_spec}Average spectra of AM\,Her at orbital
maximum (top) and orbital minimum (bottom). The top and bottom
panels give identifications of the major interstellar absorption lines
and of the emission lines, respectively.  The tick marks of the line
identifications are set to the rest wavelengths.}
\end{center}
\end{figure*}

\subsection{Average spectra}
The high spectral resolution and signal-to-noise (S/N) of the GHRS
data reveals a multitude of hitherto unexplored details in the UV
emission of AM\,Her.
Two average spectra are shown in Fig.\,\ref{ave_spec}. They contain
the typical emission lines of a polar in a state of high accretion
rate, i.e. strong emission of the high excitation lines \Ion{N}{V} and
\Ion{Si}{IV} along with weaker emission of lower ionization species,
such as \Ion{Si}{II,\,III} and \Ion{C}{II,\,III}.
The \Ion{N}{V} and \Ion{Si}{IV} doublets are fully resolved,
displaying substructure in the form of broad and narrow components.
Interestingly, the intensity of the narrow emission of
\Ion{N}{V} varies only little as a function of the orbital phase,
while the narrow emission of \Ion{Si}{IV} vanishes during orbital
maximum. The lower ionization species, \Ion{C}{II,\,III} and
\Ion{Si}{II,\,III}, do not display narrow emission at all, except for
\Line{Si}{III}{1206.5} which shows a weak narrow emission in the faint
phase spectrum.

Overlayed on the continuum and line emission from AM\,Her are numerous
interstellar lines. The very low airglow during part of the HST orbit
allows the unmistakable detection of interstellar \La\ absorption. No
obvious photospheric metal absorption lines from the white dwarf
are visible.
Note also that the broad \La\ absorption is almost completely filled
in with emission, reminiscent of the missing \Lb\ and \Lg\ absorption
lines in the ORFEUS-I FUV spectra of AM\,Her (Raymond et al. 1995).

\begin{figure}
\begin{center}
\mbox{\epsfxsize=8.8cm\epsfbox{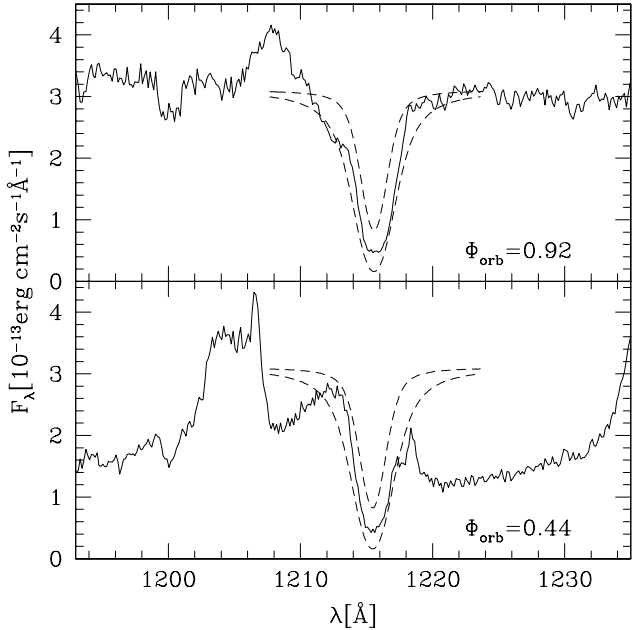}}
\caption[]{\label{islya} Interstellar \La\ absorption in GHRS spectra
of AM\,Her when HST was in the earth shadow. Orbital phases are
indicated. The dashed lines give the absorption for
$\Nhi=1\times10^{19}\rm cm^{-2}$ (upper curve) and
$\Nhi=5\times10^{19}\rm cm^{-2}$ (lower curve).}
\end{center}
\end{figure}

\subsection{Interstellar absorption lines}
During parts of the HST orbit when the satellite was in the earth
shadow, the geocoronal \La\ emission is significantly reduced, clearly
revealing interstellar \La\ absorption in the spectra of AM\,Her.
Earth-shadowed intervals occurred at $\porb\approx0.44$ and
$\porb\approx 0.92$ (Fig.\,\ref{islya}), which allow us to determine
the absorption column density during orbital minimum and maximum,
i.e. when looking either at the unheated backside of the white dwarf
or looking along the accretion funnel feeding the main pole,
respectively.
Paerels et al. (1994) found a small orbital variation of the
absorption column density from EXOSAT grating data, ranging from
$\Nh=4.2\times10^{19}\rm\,cm^{-2}$ during the soft X-ray faint phase
to $\Nh=6.7\times10^{19}\rm\,cm^{-2}$ during the soft X-ray bright
phase.  This increase of $\Nh$ is intuitively explained by the higher
mass flow rate onto the active pole.

We fitted a pure damping Lorentzian profile (Bohlin 1975) folded with
an 0.6\,\AA\ FWHM Gaussian to the observed \La\ absorption line. The
resulting {\em neutral\/} column density is
$\Nhi=(3\pm1.5)\times10^{19}\rm cm^{-2}$ for both phases, somewhat
lower than the value derived from X-ray observations,
$\Nh=(6-9)\times10^{19}\rm\,cm^{-2}$ (van Teeseling et al. 1994;
Paerels 1994; Paper\,1). There is no evidence for a phase-dependent
variation of the neutral hydrogen column density.
The higher column densities determined from X-ray data indicate the
presence of material along the line of sight, presumably within the
binary system, in which hydrogen is ionized to a high degree while the
other elements are only partially ionized and still contribute to the
soft X-ray absorption.

Assuming an average gas-to-dust ratio (Shull \& van Steenberg 1985),
our value of $\Nhi=3\times10^{19}\rm\,cm^{-2}$ translates into a limit
on the reddening of $E(B-V)\le0.006$. Considering this very low
absorption, we use throughout the following analysis the observed data
without correction for reddening.

The GHRS spectra of AM\,Her contain several interstellar absorption
lines from low ionization species in addition to \La, among which the
strongest are \Line{N}{I}{1200} (an unresolved triplet),
\Line{Si}{II}{1260.4} (blended with \Line{S}{II}{1259.5}),
\Line{O}{I}{1302.2}, \Line{Si}{II}{1304.4} and
\Line{C}{II}{1334.5}. The equivalent widths measured from the
average spectra are $\sim100$\,m\AA, comparable to those determined
for a small sample of cataclysmic variables at $d\sim100$\,pc (Mauche
et al. 1988). The centroids of the interstellar lines can be used to
check on the quality of the wavelength calibration of the GHRS
spectra, which we found to be good to $\sim30$\,\kms.

In Paper\,1, we found weak evidence for absorption of
\Lines{Si}{II}{1260.4,1265.0} in IUE low state spectra of AM\,Her,
presumably originating in the photosphere of the white dwarf. We
stress that all metal absorption lines in the GHRS spectrum are of
interstellar nature. This is underlined by the fact that only the
blue component of \Line{Si}{II}{1260.4} is detected; the red component
is caused by a transition from an excited level which is not populated
in the interstellar medium.

\begin{figure}
\begin{center}
\mbox{\epsfxsize=8.8cm\epsfbox{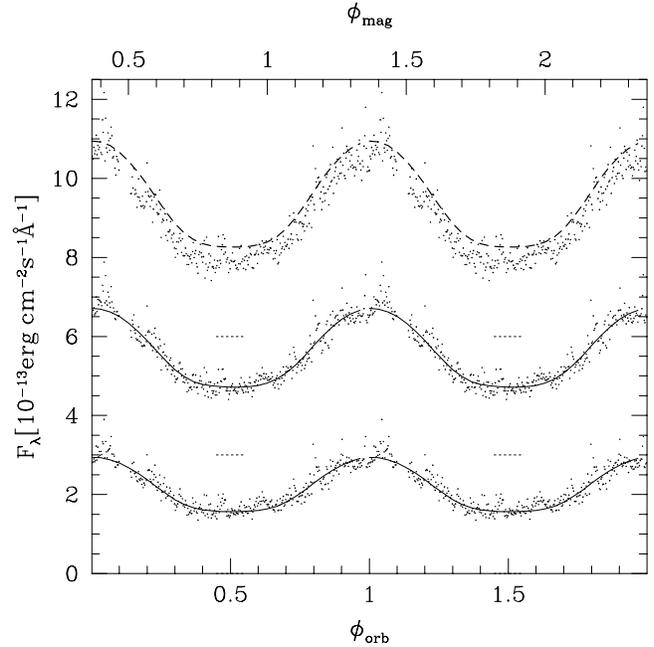}}
\caption[]{UV light curves of AM\,Her in three continuum bands. From
top to bottom: 1150$-$1167\,\AA, 1254$-$1286\,\AA, and
1412$-$1427\,\AA. The light curves are separated by 3 units in the
flux scale, dotted tick marks indicate the zero level for each
curve. The solid lines are simulated light curves from the best fit to
the observed 1254$-$1286\,\AA\ and 1412$-$1427\,\AA\ light curves. The
model flux in the 1150$-$1167\,\AA\ band (dashed line) is somewhat too
high, possibly due to remaining uncertainties in the G140L calibration, as
described in Sect.\,\ref{fitlc}.\label{lc_cont}}
\end{center}
\end{figure}

\subsection{The observed continuum flux variation}
In order to study the orbital modulation of the continuum flux, we
have selected three wavelength bands free of emission lines:
1150$-$1167\,\AA\ (Band\,1), 1254$-$1286\,\AA\ (Band\,2), and
1412$-$1427\,\AA\ (Band\,3). The continuum light curves obtained from
averaging the phase-resolved GHRS spectra in these three bands display
a quasi-sinusoidal modulation with the amplitude increasing towards
shorter wavelengths (Fig.\,\ref{lc_cont}). Maximum UV flux occurs at
$\porb\approx0.0$, which is consistent with previous IUE observations
of AM\,Her, during both the high state and the low state
(Paper\,1). The phase of the UV flux maximum agrees with that of the
maximum hard X-ray flux (e.g. Paerels et al. 1994; Paper\,1) and EUV
flux (Paerels et al. 1995), indicating that the UV excess radiation
originates close to the main accreting pole.

In Paper\,1, we ascribe the observed orbital modulation of the UV flux
to the changing aspect of a rather large, moderately hot spot on the
white dwarf. We fitted white dwarf model spectra to phase-resolved IUE
observations, obtaining a flux-weighted mean temperature and a mean
source radius from each spectrum. During the low state, the flux
variation is accompanied by a variation of the \La\ absorption
profile width, which allows a reliable determination of the
white dwarf temperature, $\Twd\simeq20\,000$\,K, and a good estimate
for the spot temperature, $\Tspot\simeq24\,000$\,K, and the spot size,
$f\simeq0.1$ of the total white dwarf surface.
However, during the high state, the broad geocoronal \La\ emission and
the unresolved \Ion{N}{V} profile are significantly blended in the
\La\ region of the IUE spectra. The derived spot temperature and size,
$\Tspot\ga37\,000$\,K and $f\la0.1$, remained, therefore, somewhat
uncertain. The main conclusion of Paper\,1 was that the flux emitted
by the UV spot roughly equals the sum of thermal bremsstrahlung and
cyclotron radiation, indicating that irradiation by the hot plasma in
the accretion column may cause the heating of the large UV spot.

\begin{figure}
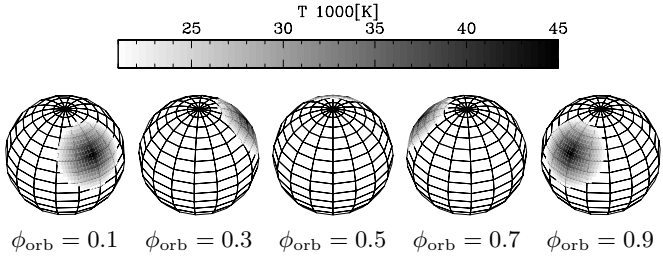

\begin{center}
\mbox{\epsfxsize6.5cm\epsfbox{f7689.04a}}

\medskip
\mbox{\epsfxsize1.68cm\epsfbox{f7689.04b}}\hfill
\mbox{\epsfxsize1.68cm\epsfbox{f7689.04c}}\hfill
\mbox{\epsfxsize1.68cm\epsfbox{f7689.04d}}\hfill
\mbox{\epsfxsize1.68cm\epsfbox{f7689.04e}}\hfill
\mbox{\epsfxsize1.68cm\epsfbox{f7689.04f}}

\parbox{1.68cm}{\centering$\porb=0.1$}\hfill
\parbox{1.68cm}{\centering$\porb=0.3$}\hfill
\parbox{1.68cm}{\centering$\porb=0.5$}\hfill
\parbox{1.68cm}{\centering$\porb=0.7$}\hfill
\parbox{1.68cm}{\centering$\porb=0.9$}
\caption[]{\label{tempmaps}Temperature maps of the best-fit model.}
\end{center}
\end{figure}

\subsection{Simulated phase-resolved spectra and light curves
\label{s-lcfit}}
In order to constrain the temperature distribution over the white dwarf
surface, we have developed a 3D white dwarf model which allows the
simulation of phase-resolved spectra and light curves.
The white dwarf surface is defined by a fine grid of several thousand
elements of roughly equal area. Each surface element is assigned an
effective temperature, allowing to prescribe an arbitrary temperature
distribution over the white dwarf surface.
For each surface element, a synthetic white dwarf spectrum with a
corresponding effective temperature is selected from a library of
model spectra computed with the atmosphere code described in Paper\,1.
Simulated phase-resolved spectra are obtained by rotating the white
dwarf model and by integrating the flux at each wavelength over the
visible hemisphere at a given phase.
The main characteristics of the model spectra are (a) pure hydrogen
composition, (b) no magnetic field, (c) $\log g=8$, and (d) no
irradiation.

The strong magnetic field of the white dwarf in polars is expected to
prevent spreading of the accreted metal-rich material perpendicular to
the field lines until the gas pressure equals the magnetic pressure, a
condition which occurs far below the photosphere. In fact, there is no
compelling observational evidence for the presence of heavy elements
in the photosphere of the white dwarf in AM\,Her (Paper\,1), so
assumption (a) therefore appears justified\footnote{ This is in strong
contrast to the situation in dwarf novae, where high metal abundances
are derived from UV spectroscopy for the photospheres of the
non-magnetic accreting white dwarfs (e.g. Long et al. 1993; Sion et
al. 1995).}.
As long as only the continuum is used for fitting purposes, the weak
Zeeman splitting of \La\ for $B\simeq14$\,MG (S.\,Jordan, private
communication) and the change of the \La\ absorption profile with
$\log g$ will not significantly affect the results. The critical point
is (d): we assume that the spectrum of a white dwarf heated by
irradiation resembles that of a somewhat hotter but undisturbed white
dwarf.
In what follows, we describe a simple model which accounts for the
main characteristics of the data.

\begin{figure}
\begin{center}
\mbox{\epsfxsize=8.8cm\epsfbox{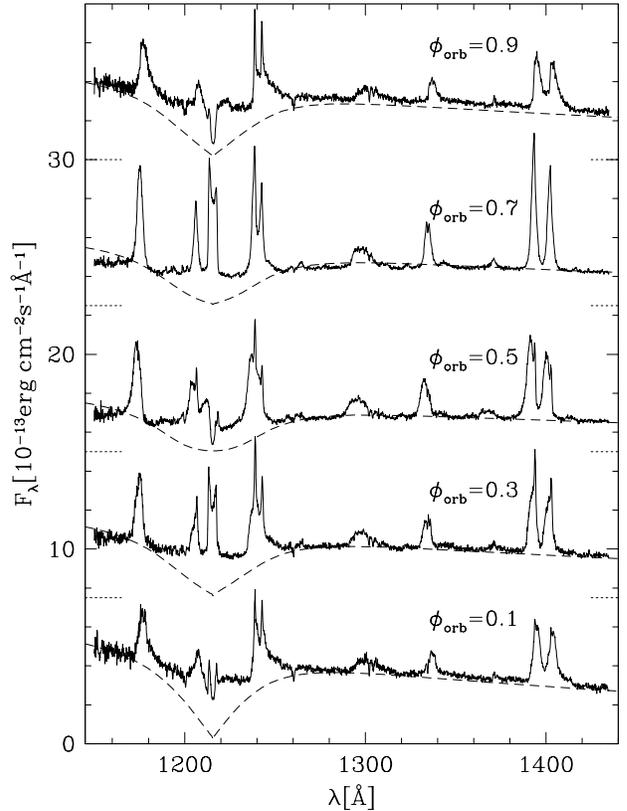}}
\caption[]{\label{pha_spec}A sample of $\Delta\phi=0.1$ phase-resolved
GHRS spectra of AM\,Her. The spectra are separated by 7.5 units in the
flux scale, dotted tick marks indicate the zero level for each
individual spectrum. Plotted dashed are simulated phase-resolved
spectra from the best-fit model. The geometry for each phase is shown
in Fig.\,\ref{tempmaps}}
\end{center}
\end{figure}

For the sake of simplicity, we chose a circular spot with an opening
angle \Oang. The spot has a temperature distribution decreasing
linearly in angle from the central value \Tcent\ at the spot centre
until meeting the temperature of the underlying white dwarf \Twd\ at
\Oang. The centre of the spot is offset from the rotational axis by an
angle \Beta\, the colatitude. Note that \Beta\ does not necessarily
equal the colatitude of the magnetic pole, commonly designated
$\beta$. The centre of the spot is offset in azimuth by an angle
$\Psi$, measured from the line connecting the centres of the two
stars.

\begin{figure*}
\begin{center}
\mbox{\epsfxsize=14cm\epsfbox{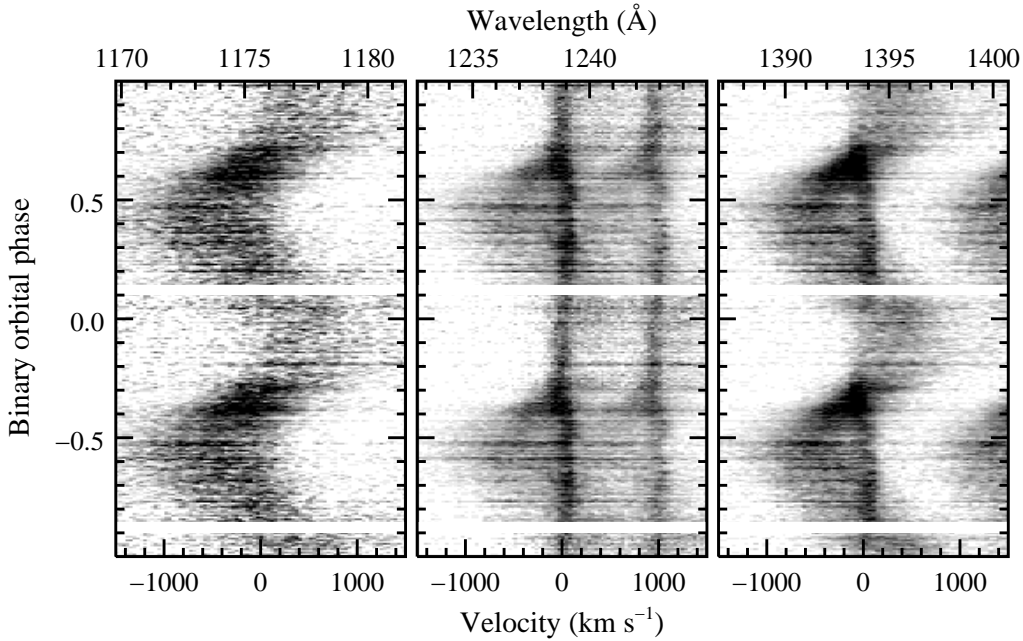}}
\caption[]{Trailed spectra of the \Line{C}{III}{1176} line
(left), \Lines{N}{V}{1239, 1243} doublet (middle), and
\Lines{Si}{IV}{1394, 1403} doublet (right) in AM Her.  The
velocity scales of the doublet lines are centred on their blue
halves, and only a small part of the red half of the \Ion{Si}{IV}
doublet is visible in the figure.\label{f-trails}}
\end{center}
\end{figure*}

\subsection{Fitting the observed light curve: Results and Caveats
\label{fitlc}}
We fitted simultaneously the observed light curves in Band\,2 and 3
only.  Because of uncertainties in the calibration of the {\em
absolute\/} flux of the G140L spectra at the very short wavelengths
(Sect.\,\ref{hstobs}), Band\,1 was not included in the fit. An
evolution strategy algorithm (Rechenberg 1994) with 6 free parameters
was used for the optimisation: the white dwarf temperature \Twd, the
white dwarf radius \Rwd, the maximum temperature of the spot \Tcent,
the opening angle of the spot \Oang, the colatitude $\beta_{\rm
spot}$, and the azimuth of the spot $\Psi$.
Fixed parameters are the distance to AM\,Her, $d=90$\,pc (Paper\,1;
Beuermann \& Weichhold 1998), and the inclination of the system,
$i=50^{\circ}$ (Davey \& Smith 1996; Wickramasinghe et
al. 1991). 

The best-fit model results in $\Twd=21\,000$\,K,
$\Rwd=1.12\times10^8$\,cm, $\Tcent=47\,000$\,K, $\Oang=69.2^{\circ}$,
$\Beta=54.4^{\circ}$ and $\Psi=0.0^{\circ}$. The opening angle
converts into a fractional spot area $f\sim0.09$. The fit proved to be
robust in \Twd, \Rwd, $\Psi$ and \Oang. However, the opening angle can
be traded for the central temperature of the spot within certain
limits (see below), because the continuum slope of the white dwarf
model spectra approaches a Rayleigh-Jeans distribution for
temperatures $\ga50\,000$\,K and becomes independent of the
temperature.

The simulated light curves in Bands\,2 and 3 from the best fit are
shown in Fig.\,\ref{lc_cont}. For completeness, Fig.\,\ref{lc_cont}
also compares the synthetic light curve in Band\,1 with the observed
one, showing that the model fluxes are somewhat too high. As noted
above, this may be due to remaining imperfections in the absolute flux
calibration of the GHRS/G140L setup at the shortest wavelengths.
The white dwarf temperature derived from the HST light curves is in
good agreement with the upper limit of 20\,000\,K based on IUE
low state spectra (Heise \& Verbunt 1988; Paper\,1). 
The same is true for the temperature and the size of the spot, even
though these are less well constrained. Smaller spots result in a
flattening of the synthetic light curve in the range
$\porb=0.37-0.57$, as the spot eventually disappears entirely behind
the limb of the white dwarf. A fit with e.g. $\Tcent=90\,000$\,K and
$\Oang=40^{\circ}$, corresponding to $f=0.03$, still yields a
satisfying result. However, it is formally not possible to exclude
even smaller spots with higher temperatures. Yet, even though not
statistically significant (see below), the observed light curve is
better fitted with a spot large enough not to be totally
self-eclipsed, as the roundness of the observed light curve in the
range $\Porb=0.35-0.65$ is better reproduced.
The temperature maps of the white dwarf surface resulting from the
best fit are shown in Fig.\,\ref{tempmaps}.

A sample of simulated phase-resolved spectra along with the GHRS
observations is shown in Fig.\,\ref{pha_spec}, clearly revealing the
major shortcoming of our approach: the model predicts a deep and broad
absorption line of \La, but the GHRS observations show only a
rather shallow \La\ absorption. The most likely cause of this
disagreement will be discussed in Sect.\,\ref{heating}.

On close inspection, the observed light curves show a weak depression
at $\porb\approx0.97$. With an inclination $i=50^{\circ}$ and a
colatitude of the spot centre $\Beta\approx55^{\circ}$, the line of
sight is almost parallel to the accretion funnel at $\porb\approx1.0$,
and part of the hot spot will be obscured by the magnetic coupled part
of the accretion stream. The weak absorption dip in the observed light
curves may be due to this shadowing effect.

There are a number of caveats concerning our model for the UV light
curve:

\noindent
(a) A proper statistical assessment of the goodness of the fit is
difficult: The statistical errors for each individual point (the
standard deviation of the flux bins in the band divided by the square
root of the number of bins) is very small. Hence, the scatter in the
observed light curves (Fig.\,\ref{lc_cont}) has some underlying
physical origin. This is not too surprising, as AM\,Her shows strong
flickering on various timescales both at optical wavelengths
(e.g. Panek 1980) and at X-rays (e.g. Szkody et al. 1980). We computed
the mean scatter in the observed light curve by subtracting a light
curve smoothed with a 30 point boxcar. Adopting the standard deviation
of this mean scatter, which is $8-10\%$ in the different bands, as the
mean error of the individual points, our best fit yields a reduced
$\chi^2=571/676$. Even though this is formally  satisfying, a
number of systematic uncertainties remain.

\begin{figure}
\begin{center}
\mbox{\epsfxsize=8.8cm\epsfbox{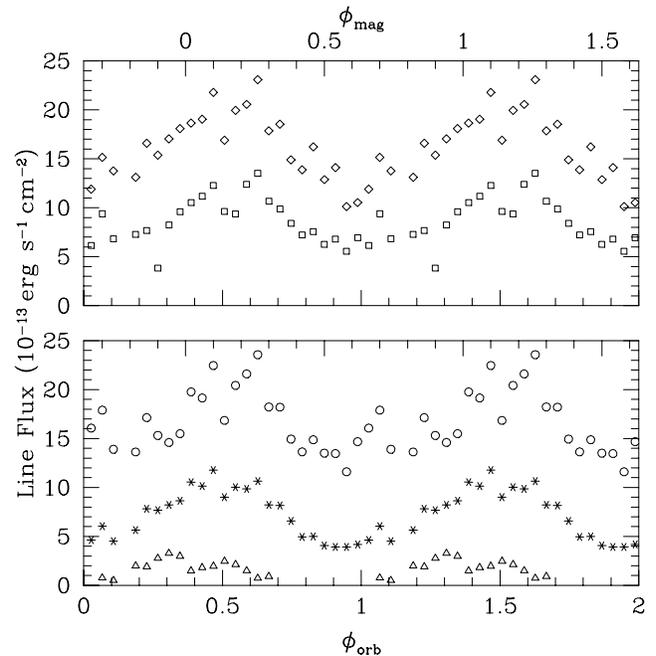}}
\caption[]{The lower panel shows emission line fluxes for \Ion{C}{II}
(asterisks), the narrow component of \Ion{Si}{IV} (triangles), and the
combined broad and high velocity components of \Ion{Si}{IV} (circles).
The upper panel shows the line fluxes for \Ion{Si}{III} (squares) and
\Ion{C}{III} (diamonds).\label{f-flux1}}
\end{center}
\end{figure}

\noindent
(b) We assume in our model that {\em all\/} the continuum flux
originates from the heated and unheated surface of the white
dwarf. This overestimates the white dwarf flux and, hence, its radius,
as the illuminated accretion stream certainly contributes to the
observed continuum flux.
In Paper\,1, we were able to quantify this continuum contribution
thanks to the broad wavelength coverage of the IUE SWP and LWP
cameras. The narrow wavelength band covered in the GHRS spectra does
not allow a similar treatment, and we rely on the results from
Paper\,1, i.e. that the contribution of the accretion stream to the
observed continuum is likely to be $\la10\%$ for
$\lambda\la1400$\,\AA.
The light curves of the broad emission lines are shifted
$\sim180^{\circ}$ in phase with respect to the continuum flux
(Sect.\,\ref{s-gaussians}; Fig.\,\ref{f-flux1}), with maximum line
flux occurring at $\porb\approx0.5$.  Therefore, the error introduced
by neglecting the stream continuum contribution is largest during the
faint phase, partially explaining the shallowness of the broad \La\
absorption at $\porb\approx0.5$.

\noindent
(c) The spot may not be circular. Modelling phase-resolved
polarimetry, Wickramasinghe et al. (1991) find a cyclotron emitting
region elongated by $10^{\circ}$ in magnetic latitude. However, unless
the spot is significantly asymmetric, the ultraviolet light curves are
not sensitive to the actual shape of the spot.

\noindent
(d) A second active region offset by $\sim180^{\circ}$ has been
detected in polarimetry (Wickramasinghe 1991) and may contribute to
the UV continuum flux during the faint phase. This would result in an
overestimate of \Twd.

\noindent
(e) The temperature distribution undoubtedly deviates from a linear
gradient in angle. An indication that this is the case comes from the
observed EUV flux of AM\,Her. Choosing $\Tcent=265\,000$\,K, as
derived by Paerels et al. (1996), it is not possible to reproduce the
observed ultraviolet modulation without exceeding the measured EUV
flux.
Assuming a linear decrease of the temperature with angle gives an
upper limit $\Tcent\la200\,000$\,K, and a lower limit
$\Oang\ga28^{\circ}$.

\subsection{Doppler Tomography\label{s-doptom}} 
Figure \ref{f-trails} shows the trailed spectra of the
\Line{C}{III}{1176} line and the \Lines{N}{V}{1239,\,1243} and
\Lines{Si}{IV}{1394,\,1403} doublet lines constructed by combining the
341 individual AM Her spectra, extracting narrow regions around the
line centres, and subtracting the continuum by means of a third-order
polynomial fit.  In the doublet lines, the zero point of the velocity
scale was set at the central wavelength of the blue doublet peak.
Both a narrow and a broad emission component are clearly visible in
the trailed spectra of the two doublet lines, while the \Ion{C}{III}
line appears to contain only a broad component.

\begin{figure}
\begin{center}
\mbox{\epsfxsize=6.0cm\epsfbox{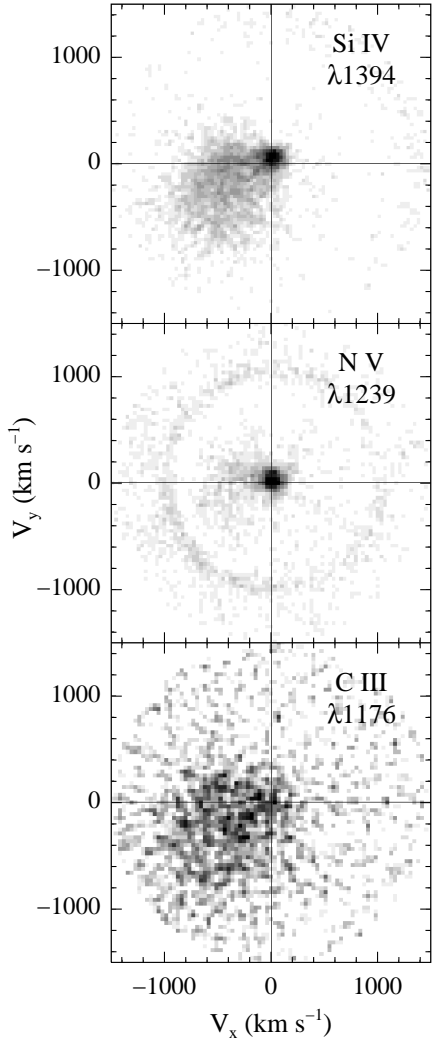}}
\caption[]{Doppler tomograms of UV lines in the spectrum of AM 
Her.\label{f-toms1}}
\end{center}
\end{figure}

The trailed spectrum of the \Ion{C}{III} line shows a broad, weak
S-wave with a full-width of $\approx1000$\,\kms\ and a maximum
(edge-to-edge) amplitude of $\approx\pm2000$\,\kms.  The line intensity
is strongest at orbital phases $\porb\approx0.6-0.7$, and weakest at
phases $\porb\approx0.8-1.1$.  Both of the doublet lines have trailed
spectra displaying a broad S-wave component similar to the
\Ion{C}{III} line, and a superimposed, narrow (full-width
$\approx100$\,\kms), low amplitude ($\la200$\,\kms) S-wave
component.  As in the \Ion{C}{III} line, the doublet line intensities
are strongest at $\porb\approx0.6-0.7$, and weakest at
$\porb\approx0.8-1.1$.  Other than the difference in blue and red
component separation, the trailed spectra of the two doublet lines
differ in one obvious respect: the narrow S-wave in the \Ion{N}{V}
line is visible throughout the entire orbit at approximately equal
strength (slightly weaker at $\porb\approx0.75-1.15$), but the narrow
component of the \Ion{Si}{IV} line completely vanishes in the phase
range 0.75$-$1.15.

In order to construct Doppler tomograms from the data, the spectra
were resampled onto a uniform velocity scale, and averaged into
orbital phase bins of width $\Delta\phi=0.01$.  This resulted in 3--4
spectra combined into each phase bin, thereby increasing the
signal-to-noise while preserving a high phase resolution.  The
tomograms were calculated using the Fourier-filtered back projection
technique (e.g., Marsh \& Horne 1988; Horne 1991).  Figure
\ref{f-toms1} shows the tomograms of the \Ion{C}{III} $\lambda1176$
line and the blue components of the \Ion{N}{V} (1239\AA) and
\Ion{Si}{IV} (1394\AA) doublets.  Tomograms for two other relatively
strong emission lines in the UV spectrum of AM Her are not shown.  The
tomogram of the \Line{C}{II}{1335} line, which lacks a narrow
component, is qualitatively similar to that of the stronger
\Ion{C}{III} line.  The trailed spectrum of the \Ion{Si}{III} emission
complex at $\approx1300$\AA\ (which also lacks a narrow component) is
qualitatively similar to that of the \Ion{C}{III} line, and we would
expect it to produce a correspondingly similar tomogram.  However, the
presence of two strong interstellar absorption features 
produced substantial artifacts in the \Ion{Si}{III} tomogram that
prevent a direct comparison with the other lines.

The ring with a radius of $\sim1000$\,\kms\ visible in the \Ion{N}{V}
tomogram is an artifact produced when the flux in the red peak of the
doublet (which is located at a velocity offset of $\approx1000$\,\kms\
from the blue peak -- see Figure \ref{f-trails}) is smeared out around
the tomogram.  A similar artifact is observed when the \Ion{Si}{IV}
tomogram is plotted to larger velocities (the doublet separation in
\Ion{Si}{IV}, $\approx2000$\,\kms, is larger than in \Ion{N}{V}).  The
red halves of the doublet lines, as should be expected, yield
tomograms identical in structure (above the noise) to the blue halves
(although the former are somewhat weaker in overall intensity than the
latter, as expected from the relative strengths of the doublet peaks,
which are also weaker on the red side).

\begin{figure}
\begin{center}
\mbox{\epsfxsize=8.8cm\epsfbox{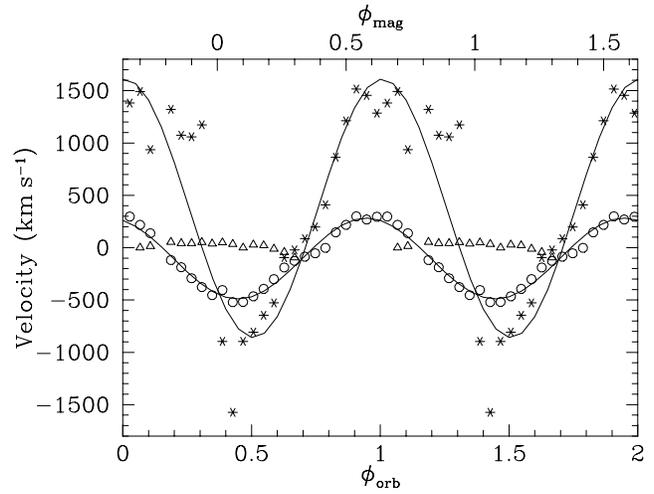}}
\caption[]{The radial velocity curve of the three emission line
components in the blue half of the \Ion{Si}{IV} doublet: the narrow
component (triangles), the broad component (circles), and the high
velocity component (asterisks).  The solid lines show the best
sinusoidal fits to the broad and high velocity
components. See Fig.\,\ref{f-vel2} for the radial velocities of the
narrow component.\label{f-vel1}}
\end{center}
\end{figure}

\begin{figure}
\begin{center}
\mbox{\epsfxsize=8.8cm\epsfbox{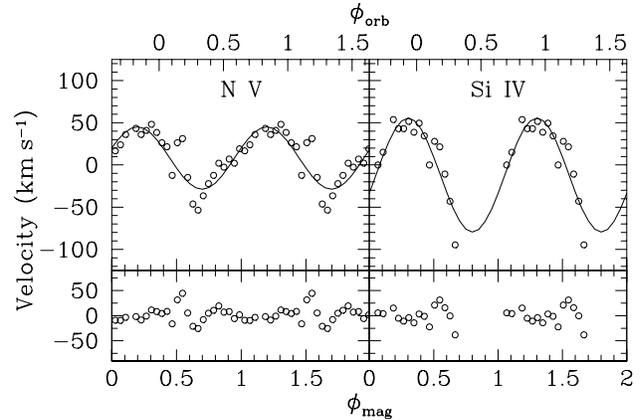}}
\caption[]{Radial velocity curves of the narrow component in the
emission lines of \Ion{N}{V} (left) and \Ion{Si}{IV} (right), along
with the best sinusoidal fits to the data (solid lines). The bottom
panels show the deviation of the data from a pure sinusoidal
shape.\label{f-vel2}}
\end{center}
\end{figure}

The \Ion{Si}{IV} tomogram contains two main emission regions: (1) a
roughly circular area of strong emission located near the velocity
origin, and (2) a broad ``fan'' of less intense emission emanating
from the velocity origin and extending (primarily) into the $(-V_{\rm
x}, -V_{\rm y})$ quadrant of the tomogram.  These emission regions can
be directly attributed to the narrow and broad emission line
components, respectively.  The general location of the strong emission
region is suggestive of an origin on or near the secondary star (see
the figure in Horne 1991 comparing velocity and position coordinates
in a CV) -- we will explore this possibility further in Section
\ref{s-narrow}.  We used contours plotted at intervals of 5\% of the
peak intensity in the tomogram (not shown) to locate the centre of the
strong emission region, at $(V_{\rm x}, V_{\rm y}) = (+10,
+90)$\,\kms\ (with an uncertainty of $\approx\pm10$\,\kms\ in each
coordinate).  This emission is asymmetrically distributed relative to
both velocity axes.  The 70\% peak contour extends from $V_{\rm
x}\approx -60$ to $+90$\,\kms\ and from $V_{\rm y}\approx -25$ to
$+135$\,\kms.

The fan-shaped emission component in the \Ion{Si}{IV} tomogram is most
likely produced by the accretion stream of AM Her.  The initial
ballistic trajectory of the stream would produce emission at
velocities starting at the $L_{1}$ point on the $+V_{\rm y}$ axis,
and extending roughly parallel to the $-V_{\rm x}$ axis for some
length before curving into the $(-V_{\rm x}, -V_{\rm y})$ quadrant
(e.g., see Horne 1991 and Figure \ref{f-toms4}), thus accounting for
the weak emission above the $-V_{\rm x}$ axis in the $(-V_{\rm x},
+V_{\rm y})$ quadrant (and, presumably, some continuing contribution
along the outer edge of the fan).  The remaining weak emission is
produced by the range of observed radial velocities of the stream
material once it is being channelled along the magnetic field lines
(e.g., Schwope et al. 1995).  As expected for a strongly magnetic CV
like AM Her, there is no ring of emission indicative of a disc in the
tomogram.  There is also no apparent enhancement to the \Ion{Si}{IV}
emission anywhere along the $-V_{\rm y}$ axis, where a contribution
from the white dwarf would be located.

\begin{figure}
\begin{center}
\mbox{\epsfxsize=8.8cm\epsfbox{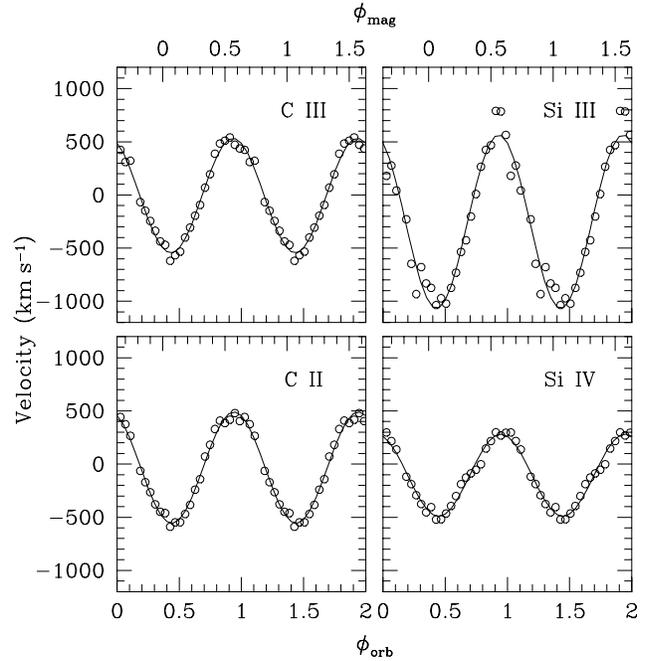}}
\caption[]{Radial velocity curves from the broad component of the
\Ion{Si}{IV} doublet and the single Gaussian fits to the \Ion{C}{II},
\Ion{Si}{III}, and \Ion{C}{III} lines, along with the best sinusoidal
fits to the data (solid lines).\label{f-vel3}}
\end{center}
\end{figure}

The \Ion{N}{V} tomogram of AM Her is similar in appearance to the
\Ion{Si}{IV} tomogram.  Again, a strong, compact emission region is
present on the $+V_{\rm y}$ axis, and a weaker fan of emission
extends to larger negative velocities.  However, the fan emission is
weaker overall than in the \Ion{Si}{IV} tomogram, and has its strongest
part along the trajectory most consistent with the initial, ballistic
accretion stream.  It is not clear if this difference in the fan
emission intensity distribution implies an actual difference in the
physical distribution of \Ion{N}{V} and \Ion{Si}{IV} emission regions
in AM Her, or if it is simply an artifact in the tomogram caused by
contamination from the broad emission component of the red half of the
more closely spaced \Ion{N}{V} doublet.

On closer inspection, we also found the strong emission region in the
\Ion{N}{V} tomogram to be somewhat different from that in the
\Ion{Si}{IV} tomogram.  Specifically, the centre of this region
(determined by plotting the 5\% intensity contours as for
\Ion{Si}{IV}) is located at $(V_{\rm x}, V_{\rm y})=(+30, +20)$\,\kms.
This implies a smaller orbital velocity around the centre-of-mass and
a higher degree of asymmetry relative to rotation around the $V_{\rm
y}$ axis than for the strong emission region in the \Ion{Si}{IV}
tomogram (see Sect.\,\ref{s-narrow}).  The 70\% contour level extends
from $V_{\rm y}\approx -50$ to $+110$\,\kms and, despite the increased
$V_{\rm x}$ offset of the emission centre, from $V_{\rm x}\approx -60$
to $+90$\,\kms (i.e., the same extent as the 70\% contour in the
\Ion{Si}{IV} tomogram).

The \Ion{C}{III} tomogram contains only a broad, fan-shaped emission
component very similar in extent to that in the \Ion{Si}{IV} tomogram.
The strong, compact emission region on the $+V_{\rm y}$ axis is
absent from this tomogram, corresponding to the lack of a narrow
component in the \Ion{C}{III} emission lines (see Fig.\ \ref{f-trails}).

\subsection{Gaussian Fits to the Emission Line Profiles\label{s-gaussians}}
In order to measure radial velocity and line flux curves for the UV
emission lines in AM Her, we binned the HST spectra into 24 orbital
phase bins of width $\Delta\phi=0.04$ (a 25th bin is empty because a
section of the CV's orbit was not observed by HST).  We then fit
single or multiple Gaussians (plus a linear function for continuum
subtraction) to the line profiles.  The number of Gaussians used for a
given line depended on the complexity of the line profile.  For the
\Ion{C}{II}, \Ion{Si}{III}, and \Ion{C}{III} lines, a single Gaussian
was used.  For the \Ion{Si}{IV} doublet, a total of 6 Gaussians were
used, 3 for each half of the doublet.  These 3 Gaussians, in turn,
were used to fit the narrow and broad line components seen in the
tomographic results (see Section \ref{s-doptom}), as well as a weak,
high velocity component revealed as a residual component during an
initial attempt to fit the \Ion{Si}{IV} doublet profiles with 2 times
2 Gaussians.  The separations of each of the 3 blue-red pairs of
Gaussians was fixed to the separation of the \Ion{Si}{IV} doublet and
their centres were allowed to vary in tandem only.  In addition, the
FWHM of the Gaussians in each of the 3 blue-red pairs were required to
be equal to each other, under the assumption that both halves of each
of the 3 doublet components should form in the same region and, hence,
display the same intrinsic velocity broadening in their profiles.  To
provide an additional constraint in fitting this complicated line
profile, the amplitudes of each blue-red Gaussian pair were required
to be related in the ratio blue:red = 1.0:0.8 (Reader et al.\ 1980).
We attempted a similar approach with the \Ion{N}{V} doublet, but the
more variable continuum and more severe blending of the doublet halves
prevented us from obtaining a reliable 6-Gaussian fit.  We were only
able to fit 2 Gaussians of fixed FWHM and separation to the narrow
components in the \Ion{N}{V} doublet in order to obtain a radial
velocity curve of the narrow component.

\begin{table}
\caption[]{Velocity curve parameters for the narrow (NC), broad (BC)
and high velocity (HVC) emission line components \label{t-vels}}
\begin{flushleft}
\begin{tabular}{lr@{$\,\pm\,$}lcrr}
\hline\noalign{\smallskip}
Line ID &
\multicolumn{2}{c}{$\gamma\,(\kms$)} &
\multicolumn{1}{c}{$K\,(\kms)$} &
\multicolumn{1}{c}{$\phi_{0}$} &
\multicolumn{1}{c}{$\sigma_{\rm tot}\FN{a}$} \\
\noalign{\smallskip}\hline\noalign{\smallskip}
\Ion{Si}{IV\,NC}  & $-12$ & 3 & $67\pm6$     & $0.55\pm0.01$ & $17$ \\
\Ion{Si}{IV\,BC}   & $-104$& 1 & $386\pm13$   & $0.20\pm0.01$ & $44$ \\
\Ion{Si}{IV\,HVC}  & $376$ &76 & $1235\pm123$ & $0.26\pm0.02$ & $416$ \\ 
\noalign{\smallskip}
\Ion{C}{II\,BC}    & $-33$ & 1 & $525\pm11$   & $0.19\pm0.01$ & $39$ \\ 
\noalign{\smallskip}
\Ion{Si}{III\,BC}  & $-251$& 2\fn{b} & $817\pm37$ & $0.18\pm0.01$ & $124$ \\ 
\noalign{\smallskip}
\Ion{N}{V\,NC}    & $10$  & 1 & $35\pm3$     & $0.47\pm0.01$ & $15$ \\ 
\noalign{\smallskip}
\Ion{C}{III\,BC}   & $-6$  & 1 & $540\pm12$   & $0.18\pm0.01$ & $44$ \\
\noalign{\smallskip}\hline
\end{tabular}
\parbox{8.8cm}{\fn{a}Total r.m.s. deviation of the data from the fit. 
\fn{b}This $\gamma$ value is not likely to be reliable since the exact
centre wavelength of the \Ion{Si}{III} emission complex is not known.
A wavelength of 1300.0\AA\ was used to calculate the velocity shifts.
}
\end{flushleft}
\end{table}

\begin{figure*}
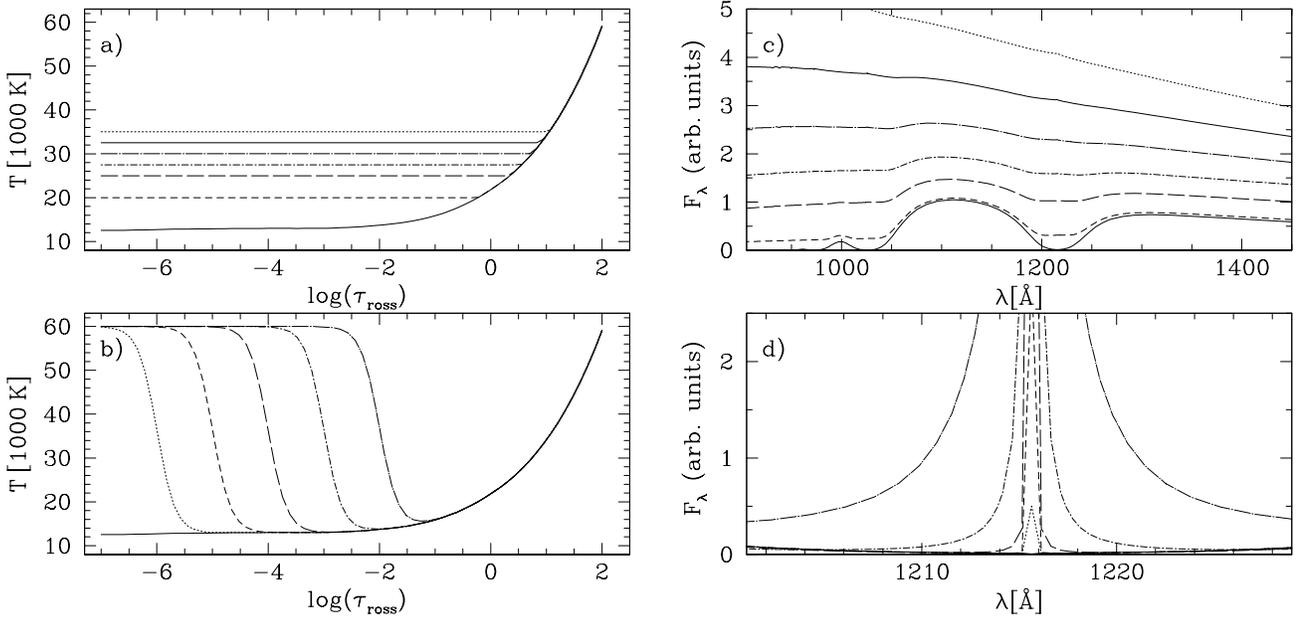

\begin{center}
\mbox{\epsfxsize=8.8cm\epsfbox{f7689.12a}
      \epsfxsize=8.8cm\epsfbox{f7689.12b}}
\mbox{\epsfxsize=8.8cm\epsfbox{f7689.12c}
      \epsfxsize=8.8cm\epsfbox{f7689.12d}}

\caption[]{``A poor man's model of the heated atmosphere''. (a) and
(b): Temperature structures modified as described in the text.  (c)
and (d): Model spectra computed from the temperature structures (a)
and (b), respectively. The line types of the individual spectra
correspond to those used for the temperature structures. The solid
curves show the temperature structure, (a) and (b), and the
spectrum, (c) and (d), of the undisturbed $\Teff=20\,000$\,K white
dwarf.\label{atmos}}
\end{center}
\end{figure*}

The centres of the Gaussians for the individual line components were
converted to velocity offsets from the corresponding line centres.  In
the case of the doublet lines, we present here only the results of the
fits to the blue halves of the doublets.  Figure \ref{f-vel1} shows
the radial velocity curves of the 3 components of the blue half of the
\Ion{Si}{IV} doublet, which are very similar to those obtained for the
3 components in the \Line{He}{II}{4686} line of the polar HU Aqr
(Schwope et al. 1997).  The best least squares fit of the sinusoidal
function
\begin{equation}
V(\porb) = \gamma - K\sin[2\pi(\porb-\phi_{0})]
\end{equation}
to the broad and high velocity components are shown in the figure.
The parameters of the fits (i.e., the systemic velocity, $\gamma$,
radial velocity semi-amplitude, $K$, and phase offset, $\phi_{0}$,
which is defined as the lag between the observed red-to-blue crossing
of the emission line velocity and the expected phase of the superior
conjunction of the white dwarf) are listed in Table \ref{t-vels}, along with
their $1\sigma$ uncertainties determined from a Monte Carlo
simulation.  The implications of the results for the narrow component
velocity curves will be discussed in Sect.\,\ref{s-narrow}.  Although
the velocity curve of the broad component is fit quite well by a sine
wave, the large parameter uncertainties and substantial $\sigma_{\rm
total}$ of the high velocity component attest to the non-sinusoidal
nature of its behaviour as a function of orbital phase.  The weak, high
velocity component detected in the \Ion{Si}{IV} line does not make a
noticeable contribution to the tomogram of that line, even when the
tomogram is plotted to larger velocities than shown in Figure
\ref{f-toms1}.  There are two main reasons for this: first, the high
velocity component is weak compared to the narrow and broad line
components, and its apparent relative brightness in the tomogram will
be further decreased by being smeared out around the high velocity
perimeter of the tomogram.  Second, the contamination from the red
half of the \Ion{Si}{IV} doublet, which occurs at a velocity offset
approximately equal to the maximum range of variability of the high
velocity component, will overpower and mask whatever contribution from
the high velocity component that might be contained in the tomogram.

Figure \ref{f-vel2} shows the velocity curves of the narrow components
of the blue halves of both the \Ion{Si}{IV} and \Ion{N}{V} doublets,
along with their best fit sine waves (see Table \ref{t-vels} for
parameter values and uncertainties).  There are no velocity points for
the narrow component of \Ion{Si}{IV} in the orbital phase interval
$\porb\approx0.7$--0.1 since the narrow component vanishes during
these phases (see Fig.\ \ref{f-trails}), probably due to obscuration
by the secondary star (Sect.\,\ref{s-narrow}).  As suggested already
by the tomograms, the implied radial velocity semi-amplitude of the
narrow component of \Ion{N}{V} is substantially smaller than that of
\Ion{Si}{IV}.  The phase offsets of $\approx0.5$ relative to superior
conjunciotn of the white dwarf for both lines imply that the source of
the narrow emission component is either on the secondary star or
follows the motion of the secondary star around the binary's centre of
mass. 

Figure \ref{f-vel3} shows the velocity curves from the broad component
of \Ion{Si}{IV} and the single Gaussian fits to \Ion{C}{II},
\Ion{Si}{III}, and \Ion{C}{III}; the fit parameters are listed in
Table \ref{t-vels}.  All four curves have phase offsets of
$\approx0.2$, but display a range of velocity semi-amplitudes:
\Ion{Si}{IV} has the smallest $K$, the two carbon lines have larger,
approximately equal $K$ values, and \Ion{Si}{III} has the largest $K$.

Figure \ref{f-flux1} shows the flux curves determined from the
Gaussian fits for the narrow component of \Ion{Si}{IV} (note that this
is the flux of the blue half of the doublet only), and for the
\Ion{Si}{III}, \Ion{C}{II}, and \Ion{C}{III} lines.  A flux curve for
the broad component of \Ion{Si}{IV} is also shown in the figure;
however, this is actually the sum of the fluxes of the broad and the
high velocity components of this line (again, the blue half of the
doublet only).  The fluxes determined individually for these
components from the Gaussian fits suffered some confusion at orbital
phases when the components' radial velocity curves crossed each other
(see Fig.\ \ref{f-vel1}).  Flux curves for \Ion{N}{V} are not
available because of the difficulty in distinguishing the blended
components of its line profile using the multi-Gaussian fitting
approach (as mentioned above).  The behaviour of the \Ion{Si}{IV}
broad component and the non-doublet line flux curves is similar in all
four of the lines, with a minimum at $\porb\approx0.9$--0.0 and a
maximum at $\porb\approx0.5$--0.6. The flux curve of the narrow
component of \Ion{Si}{IV} has a maximum at $\porb\approx0.3$, and
vanishes in the phase range $\porb\approx0.70-0.05$.

\section{Discussion}
\subsection{The temperature structure of the heated atmosphere\label{heating}}
A major difficulty in the interpretation of the origin of the UV and
EUV continuum of AM\,Her in its high state has been the absence and/or
weakness of the absorption features expected from a hot high gravity
atmosphere. The ORFEUS-I spectra of AM\,Her in high state do not show
any evidence for \Lb\ or \Lg\ absorption (Raymond et al. 1995), the
EUVE spectrum does show weak edges of \Ion{Ne}{VI,\,VIII}, but lacks
the expected strong \Ion{O}{VI\,2s,\,2p} edges (Paerels et al. 1995).
The common answer to that riddle is that heating the white dwarf by
irradiation from the post-shock plasma causes a flatter temperature
gradient in the atmosphere, weakening absorption features which form
at Rosseland optical depths $\tau_{\rm ross}<1$.
Only a limited number of irradiated white dwarf model atmospheres
exist in the literature, e.g.  Williams et al. (1987) and van
Teeseling et al. (1994). Other instructive papers, even though
treating irradiation of the secondary star in accreting binaries, are
Anderson (1981) and Brett \& Smith (1993). 
The theoretical temperature structures show mainly two features: a
thin, hot, corona-like layer at the outer boundary of the atmosphere
and a flat, sometimes completely isothermal temperature structure at
larger optical depths. For optical depths  $\tau_{\rm ross}\gg1$,
the temperature structure usually approaches that of the undisturbed
atmosphere.

A fully self-consistent model for irradiated white dwarf model
atmospheres is beyond the scope of the present paper, but we have
computed two sets of ``poor man's models'' in order to illustrate the
main observational effects that can be expected. As input, we use the
temperature structure $T(\tau)$ of an undisturbed 20\,000\,K white
dwarf, which was computed with the code described in Paper\,1. This
temperature structure was modified in two different ways.

\begin{figure}
\begin{center}
\mbox{\epsfxsize=8.8cm\epsfbox{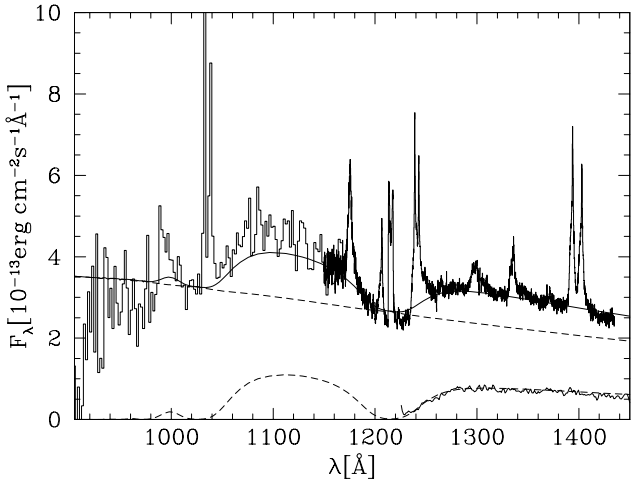}}
\caption[]{Combined ORFEUS-I (900$-$1150\,\AA) and HST/GHRS
(1150$-$1435\,\AA) spectrum of AM\,Her in high state at
$\porb=0.12-0.25$. For comparison, a low state faint-phase IUE
spectrum of AM\,Her is plotted (1225$-$1435\AA). Shown as dashed lines
are the best-fit model for the unheated backside of the white dwarf
and the contribution of the heated spot according to
Sect.\,\ref{heating}.  The solid line is the sum of the two model
components. The emission of \La\ is of geocoronal
origin. \label{hst_orf}}
\end{center}
\end{figure}

\noindent
(a) $T(\tau)$ was set to a constant value $T_{\rm out}$, from the outer
boundary of the atmosphere down to an optical depth $\tau_{\rm c}$ where
the temperature of the undisturbed white dwarf equals $T_{\rm
out}$. For optical depths larger than $\tau_{\rm c}$, $T(\tau)$ of
the undisturbed white dwarf was adopted. Figure\,\ref{atmos}a shows
the temperature structures for several values of $T_{\rm out}$. This
modification mimics an isothermal regime in the heated atmosphere.

\noindent
(b) The temperature was set to a constant $T_{\rm out}$, from the outer
boundary of the atmosphere down to an optical depth $\tau_{\rm c}$
where it is smoothly changed into the temperature run of the
undisturbed atmosphere. Figure\,\ref{atmos}b shows the temperature
structures for different values of $\tau_{\rm c}$. This modification
mimics the presence of a hot corona with a
temperature inversion.

From these modified temperature structures, we synthesised model
spectra by solving the hydrostatic equation, computing the ionization
equilibrium, occupation numbers and the absorption coefficients, and
solving the radiative transfer. The resulting spectra are shown in
Fig.\,\ref{atmos}c,d. 
In the isothermal case (Fig.\,\ref{atmos}a,c), the absorption lines
become weaker with increasing $T_{\rm out}$, and, thereby, also
increasing $\tau_{\rm c}$, and the continuum approaches the slope of a
blackbody.
If a hot layer extends deeper than to Rosseland optical depths
$\tau_{\rm ross}\simeq10^{-6}$, strong emission of \La\ is produced
(Fig.\,\ref{atmos}b,c). This is in contrast to the observations of
AM\,Her, which show no emission of \La\ during the bright phase,
i.e. when the heated region has its maximal projected area
(Fig.\,\ref{ave_spec}). During the faint phase, i.e. when the spot is
mostly eclipsed by the white dwarf, broad but weak \La\ emission is
present. This emission presumably originates in the accretion stream,
or in the outer edges of the hot spot. In any case, there is no strong
emission of
\La\ at any orbital phase in our GHRS spectra and no emission of \Lb,
\Lg\ in the ORFEUS-I spectra (Raymond et al. 1995). This indicates
that any hot corona is limited to an outermost thin layer of the
atmosphere.

With the results from our simple models in mind, we attempted a crude
two-component fit to a combined FUV/UV spectrum of AM\,Her,
constructed from the ORFEUS-I spectrum taken at $\Porb=0.12-0.25$
(Raymond et al. 1995) and GHRS data selected from the same phase
interval. Even though AM\,Her was optically fainter by 0.3\,mag during
the ORFEUS-I observations, the two spectra match quite well in
absolute flux.  Figure\,\ref{hst_orf} shows the observations along
with our two components, a model spectrum for the undisturbed white
dwarf as observed during the low state and an ``irradiated'' model
spectrum as in Fig.\,\ref{atmos}, scaled appropriately. The sum of the
two components quantitatively describes the observations for an assumed
size of the heated spot of $f\sim0.15$. This spot is uncomfortably
large, but, as discussed in Sect.\,\ref{lc_cont}, the accretion stream
might contribute somewhat to the observed UV continuum.

Thus, it appears that the heated regions of the white dwarf in AM\,Her
emit a blackbody-like spectrum without noticeable emission or
absorption features. This is in agreement with the EUV and soft X-ray
data, which gave only marginal evidence for absorption/emission
edges. 

\begin{figure}
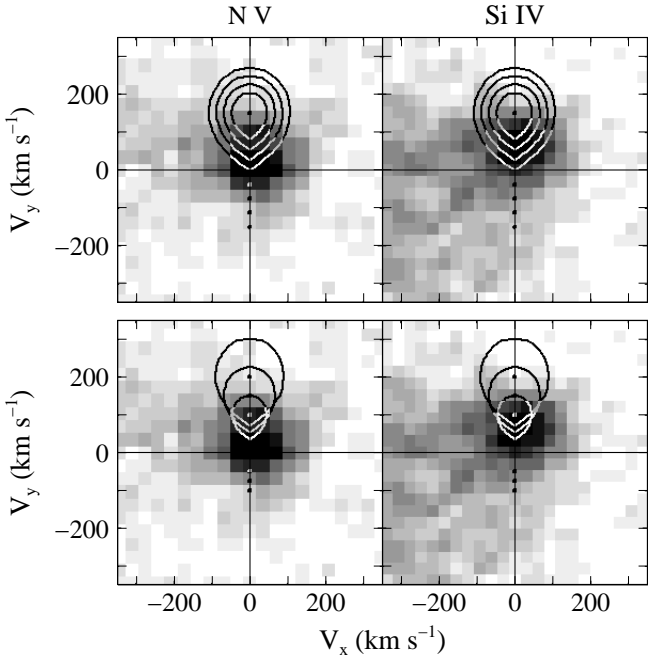

\begin{center}
\mbox{\epsfxsize=8.8cm\epsfbox{f7689.14a}}
\hfill
\mbox{\epsfxsize=8.8cm\epsfbox{f7689.14b}}
\caption[]{The central regions of the \Ion{N}{V} (left) and
\Ion{Si}{IV} (right) tomograms of AM Her, from Figure \ref{f-toms1}.
Top: Superimposed on the tomograms are the secondary star Roche lobes
for mass ratios of $q=0.25$ (innermost Roche lobe), 0.50, 0.75, 1.00
(outermost Roche lobe), assuming $K_{2}=150$\,\kms\ for all values of
$q$.  The corresponding centres of mass of the white dwarf are marked on the
$-V_{\rm y}$ axis at $K_{\rm wd}=37.5, 75, 112.5, 150$\,\kms,
respectively.\label{f-toms2}
Bottom: Secondary star Roche lobes for a constant mass ratio of
$q=0.50$ with $K_{2}=100$ (smallest Roche lobe), 150, 200 (largest
Roche lobe) \kms.  The corresponding centres of mass of the white dwarf are
marked on the $-V_{\rm y}$ axis at $K_{\rm wd}=50, 75, 100$\,\kms,
respectively.\label{f-toms3}}
\end{center}
\end{figure}

A fully satisfactory model for the phase-dependent emission of the
accretion heated white dwarf has to overcome two hurdles. (a) It is
necessary to compute self-consistent white dwarf model atmospheres
which include irradiation by thermal bremsstrahlung and cyclotron
radiation. (b) The flux and shape of the irradiating spectrum has to be
estimated as a function of the location on the white dwarf
surface. While (a) is principally a straightforward application of
model atmosphere theory, (b) includes many uncertainties with respect
to the geometry of the accretion region. The model developed by
Wickramasinghe et al. (1991) to explain polarimetric observations
could be used as a first estimate of the size and shape of the
cyclotron emitting region.

Finally, we note that the situation during the {\em low state\/} is
somewhat different: our IUE data (Paper\,1) show an almost 100\%
modulated \La\ absorption line during both the bright and faint phase.
These data can be very well fitted with model spectra of an
undisturbed white dwarf of 24\,000\,K and 20\,000\,K,
respectively. The heated side of the white dwarf, hence, appears as an
undisturbed but hotter white dwarf; the depth of the \La\ absorption
prohibits ascribing the flux modulation observed during the low state
to a heated component as described above (Fig.\,\ref{atmos}). An HST
observation of AM\,Her during a low state  is necessary order to
confirm these IUE results at a higher orbital phase resolution and a
better S/N.

\begin{figure}
\begin{center}
\mbox{\epsfxsize=8.8cm\epsfbox{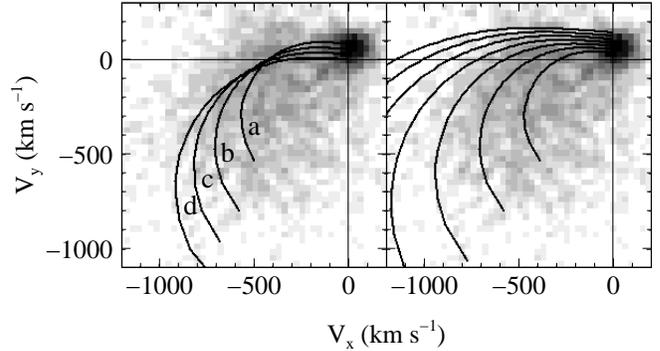}}
\caption[]{Detail of the emission regions in the \Ion{Si}{IV} tomogram
of AM Her from Figure \ref{f-toms1}.  The left panel shows the
ballistic trajectories of the accretion stream (i.e., in the absence
of a white dwarf magnetic field) for the cases discussed in the caption to
Figure \ref{f-toms2}, for $K_{2}=150$\,\kms\ and $q=$ (a) 0.25,
(b) 0.50, (c) 0.75, (d) 1.00.  The right panel shows the ballistic
trajectories of the accretion stream for a constant mass ratio of
$q=0.50$ with (starting from the innermost curve) $K_{2}=100, 150,
200, 250, 300, 350, 400$\,\kms.\label{f-toms4}}
\end{center}
\end{figure}

\subsection{Binary parameters\label{binpar}}
Various estimates for the mass of the white dwarf in AM\,Her have been
published so far, an incomplete list includes $\Mwd=0.39\,\Msun$
(Young \& Schneider 1981), $\Mwd=0.69\,\Msun$ (Wu et al. 1995),
$\Mwd=0.75\,\Msun$ (Mukai \& Charles 1987), $\Mwd=0.91\,\Msun$
(Mouchet 1993), and, the latest value, $\Mwd=1.22\,\Msun$ (Cropper et
al. 1998).
Accepting the distance to be $d\simeq90$\,pc (Paper\,1; Beuermann \&
Weichhold 1998), the UV observations constrain the white dwarf
mass. From our fit to the GHRS light curve, we obtain
$\Rwd=1.12\times10^9$\,cm, which corresponds to $\Mwd=0.35\,\Msun$,
(carbon core model; Hamada \& Salpeter 1961). This can be considered a
lower limit of \Mwd, as we assumed that all the continuum light comes
from the (heated) white dwarf; any contribution from the accretion
stream will decrease the white dwarf radius and increase its mass. On
the other hand, the flux of the IUE low state data require
$\Rwd\simeq9\times10^8$\,cm, or $\Mwd=0.53\,\Msun$. A mass as high as
1.22\,\Mwd, or $\Rwd=3.9\times10^8$\,cm, would reduce the distance to
the system to $d\approx45$\,pc in order to produce the observed UV
flux. A distance that low can be excluded both from the spectrum of
the secondary star (Paper\,1; Beuermann \& Weichhold 1998) and from
the parallax (Dahn et al. 1982). We conclude that the mass of the
white dwarf in AM\,Her is $\Mwd=0.35-0.53\,\Msun$, unless the distance
differs largely from $d=90$\,pc.

The mass of the secondary star has been estimated by Southwell et
al. (1995) to be $\Msec=0.20-0.26\,\Msun$, a result confirmed by
Beuermann \& Weichhold (1998). Thus, the range for the mass ratio
$q=\Msec/\Mwd$ is 0.38 to 0.74.

\subsection{The origin of the narrow emission lines\label{s-narrow}}
The narrow component seen in the emission lines of a number of AM Her
systems is commonly attributed to emission from the side of the
secondary star facing the white dwarf (e.g. Liebert \& Stockman
1985).  This region is heated by irradiation from the white dwarf
and/or the hot spot(s) on the white dwarf surface at the impact
point(s) of the magnetically-controlled accretion flow.
The surface of the secondary star, which is defined by the geometry of
its Roche lobe for a given mass ratio and orbital period, is preserved
in the mapping from spatial coordinates to the velocity coordinates of
a Doppler tomogram.  The secondary star's Roche lobe is symmetric about
the $+V_{\rm y}$ axis in a tomogram, with its centre offset from
$V_{\rm y}=0$ by an amount equal to the radial velocity semi-amplitude
of the secondary star, $K_{2}$.
The asymmetry relative to the $V_{\rm y}$ axis of the strong emission
regions in the \Ion{Si}{IV} and \Ion{N}{V} tomograms of AM Her
(Sect.\,\ref{s-doptom}) can be understood as due to non-uniform
heating of the secondary star's face (Smith 1995).  The accretion
stream shades the leading side of the secondary star from irradiation
by the white dwarf/hot spot(s).  This results in less heating on the
$-V_{\rm x}$ side of the $V_{\rm y}$ axis, and, therefore, stronger
emission on the $+V_{\rm x}$ side of the $V_{\rm y}$ axis. This
shielding effect is observed also in other AM\,Her stars, e.g. in
HU\,Aqr (Schwope et al. 1997).

Closer inspection of the tomograms of \Ion{N}{V} and \Ion{Si}{IV}
reveals, however, some difficulties in ascribing the entire narrow
emission to the irradiated face of the secondary
star. Figure\,\ref{f-toms3} shows blow-ups of the tomograms of
\Ion{N}{V} and \Ion{Si}{IV} from Figure\,\ref{f-toms1}. The
\Ion{Si}{IV} emission falls completely within the secondary Roche lobe
for reasonable parameters, $q=0.75$ and $K_2=150$\,\kms. However, the
\Ion{N}{V} emission requires extreme values as $q=1.0$ and
$K_2<150$\,\kms. As discussed in Sect.\,\ref{binpar}, the mass ratio
is rather unlikely to be larger than 0.75. A number of radial velocity
measurements have been obtained from the \Lines{Na}{I}{8183,\,8195}
doublet, the most reliable measurement yields $K_2=198\pm3$\,\kms\
(Southwell et al. 1995). As discussed by Southwell et al., this value
is an upper limit to the real $K$-velocity of the secondary, as the
irradiated face of the secondary contributes less to the \Ion{Na}{I}
absorption.  After some corrections, they give as a best estimate
$K_2=174$\,\kms. Considering these limits on $q$ and $K_2$, our
\Ion{N}{V} tomogram indicates narrow emission originating from
material of low velocity dispersion {\em inside\/} the Roche lobe of
the white dwarf, close to the $L_1$ point.

We can even {\em exclude\/} that significant parts of the narrow
\Ion{N}{V} emission originate on the secondary. The \Ion{Si}{IV}
narrow emission disappears at $\porb\approx0.70-1.05$
(Fig.\,\ref{f-trails}; Fig.\,\ref{f-flux1}), which is due to the
self-eclipse of the $L_1$ region on the secondary, where heating by
irradiation from the white dwarf is strongest, and where, hence,
emission is expected to be strongest. From Fig.\,\ref{f-trails}, it is
apparent that the \Ion{N}{V} narrow emission intensity does not show a
significant orbital modulation.  A consistent origin of \Ion{N}{V} on
the secondary would require the emission region to be located further
away from $L_1$ than the \Ion{Si}{IV} emission region, hence at {\em
larger\/} radial velocities. However, the \Ion{N}{V} narrow emission
has a {\em lower\/} radial velocity than the \Ion{Si}{IV} narrow
emission (Table\,\ref{t-vels}).

We conclude that the GHRS spectra clearly demonstrate the presence of
highly ionized low-velocity-dispersion material which co-rotates with
the binary and is located between $L_1$ and the centre of mass. A
tempting possibility is that this material is kept in place in a
magnetic slingshot prominence emanating from the secondary
star. Evidence for such prominences have been found before in the
dwarf novae IP\,Peg and SS\,Cyg (Steeghs et al. 1996). We will explore
the physical properties of the material hold in the prominence in a
future paper.

\subsection{On the broad emission lines\label{s-broad}}
A consistent interpretation of the broad emission lines encounters
some difficulties.  The radial velocity curves (Fig.\,\ref{f-vel3})
clearly indicate an origin in the accretion stream, with maximum
redshift at $\porb\approx0.9-1.0$, i.e. when looking parallel to the
stream, and maximum blueshift at $\porb\approx0.4-0.5$, i.e. when
looking anti-parallel to the stream.
The Doppler tomograms (Fig.\,\ref{f-toms4}) show that most emission is
centred around the ballistic part of the accretion stream if
reasonable values are chosen for the mass ratio, e.g. $q=0.5$ and
$K_2=150-200$\,\kms.

The single-humped light curves of the broad emission lines with
maximum flux at $\porb\approx0.5$ and minimum flux at
$\porb\approx0.9$ (Fig.\,\ref{f-flux1}) are not quantitatively
understood.
On one hand, if the accretion stream were optically thick, one would
naively expect a double-humped light curve: the projected area of the
stream is minimal at phases $\porb\approx0.0,0.5$ and maximal during
phases $\porb\approx0.25,0.75$. This behaviour is observed e.g. in
HU\,Aqr (Schwope et al. 1997) and QQ\,Vul (G\"ansicke 1997).
On the other hand, if the stream were optically thin, one would expect
no variation of the line fluxes around the binary orbit.
Optical depth effects, self-eclipse and irradiation of the stream may
be responsible for the observed variation of the broad line fluxes.

An interesting result of the GHRS observations is the detection of a
high-velocity component in the emission of \Ion{Si}{IV}. The low flux
of this component, however, prevents a location of its origin in the
Doppler tomograms, as already noted in Sect.\,\ref{s-doptom}. The
similar phasing of the radial velocity curves of the broad and the
high velocity component (Fig.\,\ref{f-vel1}) indicates that both
components originate in regions of the accretion stream not too far
apart, with the source of the high-velocity emission probably
being located closer to the white dwarf.

\section{Conclusion}
Phase-resolved HST/GHRS observations of AM\,Her in a high state have
revealed a wealth of previously unobserved details.

\begin{enumerate}
\item We measure a column density
$\Nhi=(3\pm1.5)\times10^{19}\rm\,cm^{-2}$ from the interstellar \La\
absorption profile. There is no evidence for a phase-dependence of the
absorption column. The present high state spectra also show no
evidence for photospheric absorption lines from the white
dwarf atmosphere other than shallow \La.
\item We confirm our previous findings that the UV continuum flux
modulation is due to a hot spot on the white dwarf and we successfully
fit the observed light curve, deriving an upper limit on the spot size
of $f\sim0.09$. However, a smaller and hotter spot can not be
excluded. 
\item The absence/weakness of strong Lyman absorption lines in our
GHRS data as well as in the ORFEUS-I data can be understood if the hot
spot radiates as a blackbody. The contribution of the unheated parts
of the white dwarf is sufficient to explain the shallow \La\
absorption in the GHRS spectra.
\item The Doppler tomograms computed from the emission lines show
that, while the narrow emission of \Ion{Si}{IV} originates from the
heated face of the secondary, the narrow emission of \Ion{N}{V} must
originate from material between $L_1$ and the centre of mass. A
promising way to keep material co-rotating with the binary is a
magnetic prominence located on the secondary star near the $L_1$
point.
\item For a distance $d=90$\,pc, the mass of the white dwarf in
AM\,Her is $\Mwd=0.35-0.5$.
\end{enumerate}

\begin{acknowledgements}
We thank Chris Mauche for communicating the ORFEUS-1 spectra of
AM\,Her, Jens Kube for discussions on the accretion geometry in
AM\,Her, Stefan Jordan for discussions on magnetic white dwarf model
spectra, Janet Mattei for communicating the visual magnitudes of
AM\,Her, Rick Hessman and Axel Schwope for comments on an earlier
version of the manuscript, and Howard Lanning for his support during
the HST observations. Finally, we thank the referee Frank Verbunt for
helpful comments.
BTG was supported in part by the DARA under project number
50\,OR\,9210. PS and DWH acknowledge support from NASA HST grant
GO-06558.02-95a. Participation by EMS was supported by NASA HST grant
GO-06885.01-95a and by NSF grant AST90-16283-A01 to Villanova
University.

\end{acknowledgements}


\begin{thebibliography}{}
%
\bibitem[Anderson 1981]{and81} 
 Anderson L., 1981, ApJ 244, 555
%
\bibitem[Beuermann \& Weichhold 1998]{bw98} 
 Beuermann K., Weichhold M., 1998, A\&A {\em submitted}
%
\bibitem[Bohlin 1975]{boh75} 
 Bohlin R.C., 1975, ApJ 200, 402
%
\bibitem[Brett \& Smith 1993]{bs93} 
 Brett J.M., Smith R.C., 1993, MNRAS 264, 641
%
\bibitem[Cropper et al. 1998]{crw98} 
 Cropper M., Ramsay G., Wu K., 1998, MNRAS 293, 222
%
\bibitem[Dahn et al. 1982]{dahea82} 
 Dahn C.C., Riepe B.Y., Guetter H.H., et al., 1982, AJ 87, 419
%
\bibitem[Davey \& Smith 1996]{ds96} 
 Davey S.C., Smith R.C., 1996, MNRAS 280, 481
%
\bibitem[G\"ansicke et al. 1995]{gaeea95} 
 G\"ansicke B.T., Beuermann K., de Martino D., 1995, A\&A 303, 127 (Paper\,1)
%
\bibitem[G\"ansicke 1997]{gae97} 
 G\"ansicke B.T.,  1997, PhD dissertation, University of G\"ottingen
%
\bibitem[Greenstein et al. 1977]{greea77} 
 Greenstein, J.L., Sargent, W.L.W., Boroson, T.A., Boksenberg, A.,
 1977, ApJ 1977, L21
%
\bibitem[Hamada \& Salpeter 1961]{hs61} 
 Hamada T., Salpeter E.E., 1961, ApJ 134, 683
%
\bibitem[Heise \& Verbunt 1988]{hv88} 
 Heise J., Verbunt F., 1988, A\&A 189, 112
%
\bibitem[Horne 1991]{Hor91} %
 Horne K., 1991, in {\em Proceedings of the 12th North American
 Workshop on Cataclysmic Variables and Low Mass X-ray Binaries\/}, ed.
 A. W. Shafter (San Diego: San Diego State University), p.\,23
%
\bibitem[Liebert \& Stockman 1985]{ls85} %
 Liebert J., Stockman H.S., 1985, in {\em CVs and LMXBs\/}, eds
 D.Q. Lamb, J. Patterson, p.\,151
%
\bibitem[Long et al. 1993]{lonea93} 
 Long, K.S., Blair, W.P., Bowers, C.W., Davidson, A.F., Kriss, G.A., 
 Sion, E.M., Hubeny, I., 1993, ApJ 405, 327
%
\bibitem[Mauche et al. 1988]{mauea} 
 Mauche C.W., Raymond J.C., C\'ordova F.A., 1988, ApJ 355, 829
%
\bibitem[Marsh \& Horne 1988]{MH88} 
 Marsh T. R. \& Horne K., 1988, MNRAS, 235, 269
%
\bibitem[Martin 1988]{Mar88} 
 Martin J. S., 1988, PhD dissertation, University of Sussex
%
\bibitem[Mukai \& Charles 1987]{mc87} %
 Mukai K., Charles P.A., 1987, MNRAS 226, 209
%
\bibitem[Mouchet 1993]{mou93} 
 Mouchet M., 1993, in {\it White Dwarfs: Advances in Observation and
 Theory\/}, ed Barstow, M., p.\,411 (Dordrecht: Kluwer)
%
\bibitem[Paerels et al. 1994]{paea94} 
 Paerels F., Heise J., van Teeseling A., 1994, ApJ 426, 313
%
\bibitem[Paerels et al. 1996]{paeea96} 
 Paerels F., Young Hur M., Mauche C.W., Heise J., 1995, ApJ 464, 884
%
\bibitem[Panek 1980]{pan80} 
 Panek R.J., 1980, ApJ 241, 1077
%
\bibitem[Raymond et al. 1995]{rayea95} 
 Raymond J.C., Mauche C.W., Bowyer S., Hurwitz M., 1995, ApJ 440, 331
%
\bibitem[Reader et al. 1980]{RCW80}
 Reader J., Corliss C. H., Wiese W. L., \& Martin G. A.,
 1980, in Wavelengths and Transition Probabilities for Atoms and Atomic 
 Ions (Washington, D. C.: U. S. Goverment Printing Office), 135
%
\bibitem[Rechenberg 1994]{rec94} %
 Rechenberg I., 1994, {\sl Evolutionsstrategie '94\/}
 (froommann--holzboog: Stuttgart) 
%
\bibitem[Schwope et al. 1995]{SSM95} 
 Schwope A. D., Schwarz R., Mantel K.-H., Beuermann K.,  1995,
 in {\em Cape Workshop on Magnetic Cataclysmic Variables\/},
 eds. D.A.H. Buckley, B. Warner, p.\,166
%
\bibitem[Schwope et al. 1997]{SMH97} 
 Schwope A. D., Mantel K.-H., Horne K., 1997, A\&A, 319, 894
%
\bibitem[Sherbert \& Hulbert 1997]{sh97} 
 Sherbert L.E., Hulbert S.J., 1997, GHRS-ISR 67 update, STScI
%
\bibitem[Sherbert et al. 1997]{ssm97}
 Sherbert L.E., Soderblom D.R., Mack J., 1997, GHRS-ISR 85, STScI
%
\bibitem[Shull \& van Steenberg]{ss85}
 Shull J.M., van Steenberg M.E., 1985, ApJ 294, 599
%
\bibitem[Sion et al. 1995]{sioea95} 
 Sion, E.M., Szkody, P., Cheng, F.H., Huang, M., 1995, ApJ 444, L97
%
\bibitem[Southwell et al. 1995]{souea95} 
 Southwell K.A., Still M.D., Smith R.C., Martin J.S., 1995, A\&A
 302, 90
%
\bibitem[Smith 1995]{smi95} 
 Smith, R.C., 1995,  in {\it Cape Workshop on Magnetic
 Cataclysmic Variables\/}, eds. Buckley, D.A.H. \& Warner, B., p.\,417
 (ASP Conf. Ser. 85)
%
\bibitem[Steeghs et al. 1996]{steea} 
 Steeghs D., Horne K., Marsh T.R., Donati J.F., 1996,  MNRAS 281, 626
%
\bibitem[Szkody et al. 1980]{szkea80} 
 Szkody P., Cordova F.A., Tuohy I.R., et al., 1980, ApJ 241, 1070
%
\bibitem[van Teeseling et al. 1994]{thp94} 
 van Teeseling A., Heise J., Paerels F., 1994, A\&A 281, 119 
%
\bibitem[Warner 1995]{} 
 Warner, B., 1995, {\sl Cataclysmic Variable Stars\/} (Cambridge
 University Press)
%
\bibitem[Wickramasinge et al. 1991]{wicea91} 
 Wickramasinghe D.T., Bailey J., Meggitt S.M.A., et al., 1991, MNRAS 251, 28
%
\bibitem[Williams et al. 1987]{wilea87} 
 Williams G.A., King A.R., Brooker J.R.E., 1987, MNRAS 266, 725
%
\bibitem[Wu et al. 1995]{wcs95} 
 Wu K., Changmugam G., Shaviv G., 1995, ApJ 455, 260
%
\bibitem[Young \& Schneider 1979]{ys79} 
 Young P., Schneider D.P., 1979, ApJ 230, 502
%
\bibitem[Young \& Schneider 1981]{ys81} 
 Young P., Schneider D.P., 1981, ApJ 245, 1043
%
\end{thebibliography}
\end{document}